\pdfoutput=1
\documentclass[reprint,twocolumn,aps,prl,showpacs,superscriptaddress,citeautoscript]{revtex4-1}
\usepackage{graphicx}  
\usepackage{dcolumn}   
\usepackage{bm}        
\usepackage{amssymb}  
\usepackage{amsmath}

\newcommand{\beq}{\begin{equation}}
\newcommand{\eeq}{\end{equation}}

\DeclareMathOperator{\sech}{sech}

\begin{document}
\title{Gradient Index Devices for the Simultaneous Focusing of the $S_0$ and $A_0$ Lamb Modes}
\date{\today}

\author{Yabin Jin}
\affiliation{School of Aerospace Engineering and Applied Mechanics, Tongji University, 100 Zhangwu Road, 200092, Shanghai, China}
\affiliation{Institut d'Electronique, de Microl\'ectronique et de Nanotechnologie, UMR CNRS 8520, Universit\'e de Lille 
1, 59655 Villeneuve d’Ascq, France}

\author{Daniel Torrent}
\email{daniel.torrent@iemn.univ-lille1.fr}
\affiliation{Institut d'Electronique, de Microl\'ectronique et de Nanotechnologie, UMR CNRS 8520, Universit\'e de Lille 
1, 59655 Villeneuve d’Ascq, France}

\author{Yan Pennec}
\affiliation{Institut d'Electronique, de Microl\'ectronique et de Nanotechnologie, UMR CNRS 8520, Universit\'e de Lille 
1, 59655 Villeneuve d’Ascq, France}

\author{Yongdong Pan}
\affiliation{School of Aerospace Engineering and Applied Mechanics, Tongji University, 100 Zhangwu Road, 200092, Shanghai, China}

\author{Bahram Djafari-Rouhani}
\affiliation{Institut d'Electronique, de Microl\'ectronique et de Nanotechnologie, UMR CNRS 8520, Universit\'e de Lille 
1, 59655 Villeneuve d’Ascq, France}

\begin{abstract}
Gradient index devices for the simultaneous control of the $S_0$ and $A_0$ Lamb modes are presented and numerically analysed. It is found a special relationship between the refractive indexes for these two modes which allows their simultaneous control by means of graded phononic crystals and thickness variations of the plate. A flat gradient index lens and a circular Luneburg lens are designed. Numerical simulations show that the performance of these devices is good for the two modes in a broadband frequency region, and that this approach can be used to design more advanced refractive devices for the total control of guided vibrational modes.
\end{abstract}
\maketitle

Gradient index (GRIN) devices consist of locally inhomogeneous materials in which the refractive index is a function of the spatial coordinates, therefore waves follow curved trajectories that can be properly designed to focus the energy at a specific point. These devices, which are well known in optics\cite{vsarbort2012spherical}, are recently receiving increasing attention in the field of phononic and sonic crystals, given that these structures can be properly designed to work as GRIN devices. 

Effectively, phononic crystals are essentially periodic arrangements of solid inclusions in an elastic matrix (fluid background for sonic crystals) and they behave, in the low frequency limit, as homogeneous materials with effective constitutive parameters which depend mainly on the filling fraction of the inclusions\cite{homoArkadii}. Then, by locally changing this parameter according to some specific law, we can create an artificially inhomogeneous material which will work as a gradient index device. Following this approach, a wide variety of inhomogeneous devices have been proposed, like GRIN flat lenses \cite{GRINLin,Climente2010,Martin2010,zigoneanu2011design,martin2012elastic}, acoustic beam modifiers \cite{lin2012design}, omnidirectional refractive devices \cite{climente2014gradient,torrent2014omnidirectional} or absorbers \cite{Climente2012,umnova2012omnidirectional} and acoustic or elastic cloaks\cite{cummer2007one,norris2011elastic}, which are indeed a new type of advanced gradient index device.

In the case of GRIN devices for solids, the main applications arise in the propagation of guided waves, like Lamb or Love waves in plates and substrates\cite{royer2000elastic}.  However, these systems present several propagation modes and polarizations, being their dispersion relations different in general. The main effort so far in this domain has been focused on the design of GRIN devices for the control of the lower order Lamb mode in thin plates, i.e., for the control of flexural waves\cite{wu2011focusing,zhao2012efficient,yan2013focusing}. 

In this work, the performance of this type of devices is improved, by exploiting an interesting property of the two lower order Lamb modes, which is the fact that they are connected by means of the so called ``plate velocity'', and it is shown that it is therefore possible to design of GRIN devices for the simultaneous control of these two modes in a broadband frequency region.

The dispersion relation of Lamb waves in elastic plates can be found in reference \onlinecite{royer2000elastic}, where it is also shown that in the low frequency limit, defined by means of wavelengths $\lambda>h$, being $h$ the thickness of the plate, the dispersion relation for the $S_0$ mode is linear and given by
\beq
\label{eq:ks}
\omega=v_p k_S,
\eeq
while for the $A_0$ mode it is parabolic and given by
\beq
\label{eq:ka}
\omega=\frac{v_ph}{\sqrt{12}}k^2_A,
\eeq
being $v_p$ the so called ``plate velocity''. These expressions show the interesting property that the dispersion relation of the $A_0$ mode depends on the plate's thickness $h$, and it is directly related with the dispersion relation of the $S_0$ mode by means of the plate velocity $v_p$. In this work this relationship between the dispersion relation of the two modes will be exploited, and homogenized two-dimensional phononic crystals will be employed to that purpose. 

The above expressions are valid  for the vibrations of a homogeneous and isotropic plate, and it must be figured out if they can also be applied to phononic crystal plates in the low frequency limit. Phononic crystal plates consist of periodic arrangements of inclusions in a plate, and they can be considered like ``slides'' of a two-dimensional phononic crystal consisting of periodic arrangements of cylinders. This two-dimensional phononic crystal will behave, in the low frequency limit, like a homogeneous but anisotropic material, given that the properties in the $xy$ plane are different from those in the z direction. Then, a ``slide'' of this phononic crystal will not vibrate like a homogeneous and isotropic plate, but like a homogeneous and anisotropic plate. It is not clear therefore that expressions \eqref{eq:ks} and \eqref{eq:ka} be valid in this case. 

The analysis of Lamb waves in anisotropic plates is complex, but in the specific case of tetragonal materials (triangular or square arrangement of inclusions), which is the case considered here, it can be demonstrated (see supplementary material) that expressions \eqref{eq:ks} and \eqref{eq:ka} are valid but the plate velocity must be given by
\beq
v_p^2=\frac{c_{11}}{\rho}\left(1-\frac{c_{13}^2}{c_{11}c_{33}}\right),
\eeq
where $c_{ij}$ stands for the $ij$ element of the stiffness tensor of the plate. 

The propagation of waves in a medium is determined by the dispersion relation, but the refraction of waves when passing from one medium to another one is managed by the ratio between the wavenumbers in the two media, which defines the index of refraction. Then, it is easy to check that if a wave changes from propagation medium 1 to medium 2, the refractive index for the $S_0$ mode will be
\beq
\label{eq:nvp}
n_S=\frac{v_{p1}}{v_{p2}},
\eeq
while for the $A_0$ mode will be
\beq
\label{eq:nans}
n_A^2=n_S\frac{h_1}{h_2},
\eeq
where it has been assumed that the two media are two different plates, with different materials and thickness. The above two expressions show that, although the dispersion relation for the two modes is different, with different frequency-dependence, their refractive indexes are closely related, with the remarkable result that, if the thickness of the two media is chosen such that $h_1=h_2n_S$, it will be found that the refractive index of the $S_0$ and the $A_0$ mode is the same. This interesting property will be exploited here for the design of a refractive device identical for the two modes. 

Two devices are proposed in this work based on the combination of graded phononic crystals and thickness variations. The first one consists in a gradient index lens and the second one consists in a Luneburg lens. These devices have been already studied for flexural waves, and here it is proposed their design to work simultaneously for the two fundamental Lamb modes. 

A GRIN lens consists in a rectangular device where the refractive index is a function of the distance to the lens' axis, being such a function given by\cite{GRINLin}
\beq
\label{eq:ny}
n(y)=n_0\sech(\alpha y),
\eeq
so that if the gradient of the index is along the $y$ direction, a wave travelling along the $x$ axis and arriving to the device is focused at a distance $f=\pi/2\alpha$. Similarly, a Luneburg lens consist of a circular device in which the refractive index is a function of the distance to the center of the lens and given by\cite{vsarbort2012spherical}
\beq
n(r)=\sqrt{2-(r/R_L)^2},
\eeq
where $R_L$ is the radius of the lens. This device is an omnidirectional device in which any plane wave arriving at its surface is focused on the diametrically opposed border of the lens.

Phononic crystal based GRIN devices consist in the realization of the above variations of the refractive index by means of a variation in the physical properties of the inclusions. In the low frequency limit phononic crystals behave like homogeneous materials with some effective parameters which are function of the inclusions' properties and sizes, then a rectangular or circular region of a phononic crystal can be properly designed to satisfy the above position dependent refractive indexes. 

The design of this type of devices requires therefore of a proper effective medium theory which gives the effective refractive index of the phononic crystal as a function of the physical properties of the inclusions. In the present case, this theory is the recently developed homogenization theory for flexural waves ($A_0$ mode) which can be used to obtain the refractive index of the $A_0$ mode \cite{torrent2014effective}. In the framework of this theory the propagation of waves is described in terms of the rigidity $D_b$ of the plate and its mass density $\rho_b$, and if in the plate there is a phononic crystal, the theory gives the effective rigidity $\bar{D}^*$ and mass density $\bar{\rho}^*$ relative to that of the background $D_b$ and $\rho_b$. In the supplementary material it is shown that the plate's velocity in the phononic crystal region is related with this theory by 
\beq
\label{eq:vpDrho}
\bar{v}_p^ 2=\frac{\bar{D}^*}{\bar{\rho}^*}\frac{E_b}{\rho_b(1-\nu_b^2)}
\eeq
so that from the homogenization theory for flexural waves it is possible to deduce the plate's velocity and, consequently, the refractive index not only for the $A_0$ mode but also for the $S_0$ mode.

The design procedure for the GRIN lens is therefore as follows. First, an inclusion located at $\bm{R}_\alpha=(x_\alpha,y_\alpha)$ is selected, the expected refractive index at that position is computed by means of equation \eqref{eq:ny} and by means of equation \eqref{eq:nvp} it is obtained the corresponding plate velocity. Equation \eqref{eq:vpDrho} relates the plate velocity with the properties of the inclusion, since the objective is to create the GRIN lens by changing the filling fraction, from this equation we solve for the expected filling fraction and, subsequently, for the inclusion's radius. The procedure is repeated to all the inclusions in the plate so that the design of the full GRIN lens is finally completed. Notice that it has been employed the effective medium theory for flexural waves to obtain the effective parameters, but equation \eqref{eq:nvp} for the design of the lens, what means that the GRIN designed works only for the $S_0$ mode. 

The key point in this work is that it is possible to choose now a position dependent thickness of the plate such that $h_2(y)=h_b/n(y)$, being $h_b$ the thickness in the background (i.e. outside the lens region), given that $n(y)=n_S(y)$ and following equation \eqref{eq:nans}, it will be found that the refractive index of the two modes is identical and, therefore, the GRIN lens will work properly for the two modes.

The designed lens consists of a square arrangement with lattice constant $a$ of lead inclusions in a silicon plate of thickness $h=a$, with $\rho_a=11.34$ kg/m$^3$, $E_a=16$ GPa and $\nu_a=0.44$ the mass density, Young's modulus and Poisson's ratio for lead and $\rho_b=2.33$ kg/m$^3$, $E_b=150$ GPa and $\nu_b=0.28$ the corresponding parameters for silicon. In the low frequency limit this arrangement of inclusions will behave like a tetragonal material, but it has been mentioned before and demonstrated in the supplementary material that the basic relationships used here work as well for this type of plates. The rectangular region of the phononic crystal is selected as having $N_y=15$ rows and $N_x=30$ columns, which give us a lens of height $L_y=15a$ and length $L_x=30a$.


\begin{figure}
\centering
\includegraphics[width=\columnwidth]{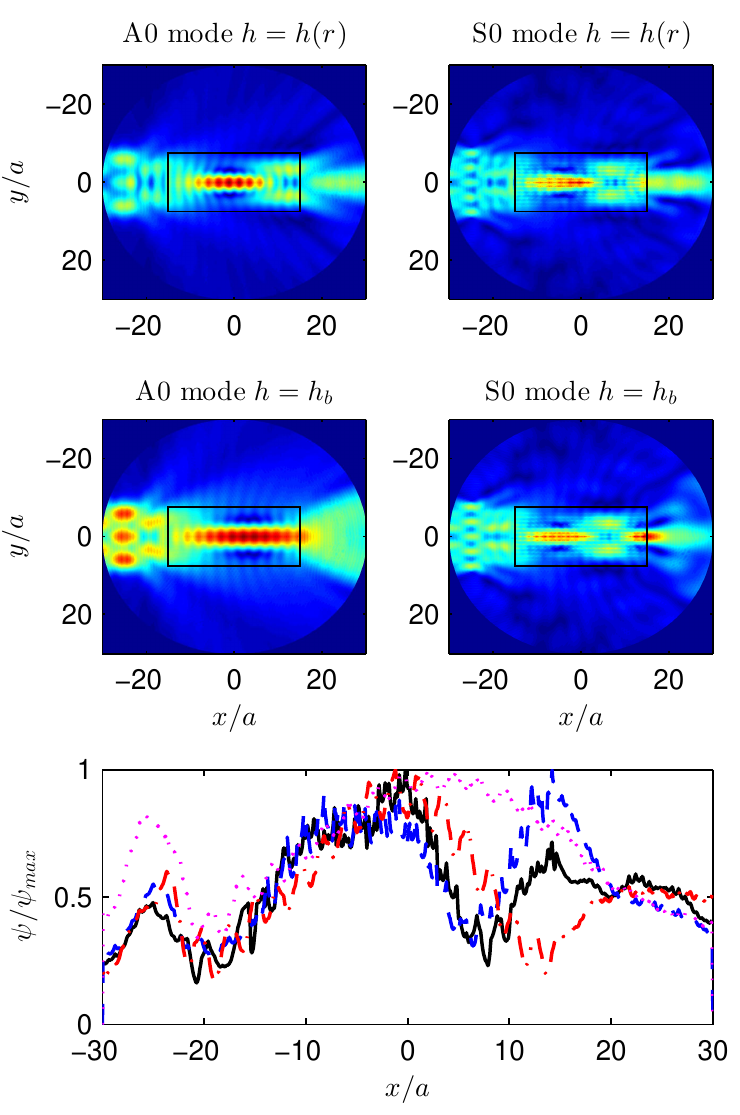}
\caption{\label{Fig:field_knumber} Upper Panels: Field distribution for $ka=1$ for a $A_0$ polarized (left) and a $S_0$ (right) polarized plane waves interacting with the gradient index lens where the thickness of the plate is designed for the simultaneous focusing of the two Lamb modes. Mid panels: Same as the upper panel but now the thickness of the plate is constant throughout the lens, so that the focusing of the $A_0$ mode is different than for the $S_0$ mode. Lower panel:Normalized field distribution along the $x$ axis and for $y=0$ (central axis of the lens) for the $S_0$ mode with thickness variation (black continuous line) and without thickness variation (blue dashed line), and for the $A_0$ mode with thickness variation (red dash-doted line) and without thickness variation (magenta doted line). It is seen that the thickness variation does not affect the $S_0$ mode, while for the $A_0$ mode it changes the focusing point, making it identical to that of the $S_0$ mode}
\end{figure}

Figure \ref{Fig:field_knumber} shows COMSOL simulations of the interaction of the GRIN lens with the $S_0$ and $A_0$ polarizations. Upper panels show the proper GRIN design in which the thickness of the plate in the rectangular phononic crystal region is position-dependent, while mid panels show the GRIN lens without the mentioned thickness variation. Left panels correspond to the the normalized $z$ component of the displacement field for the $A_0$, and the right ones to the $S_0$ mode, in which it is plotted the normalized $x$ component of the displacement field. Given that the performance of the device depends on the wavelength of the field and not on the frequency, simulations are shown for frequencies in which the two modes have the same wavenumber $ka=1$, which corresponds to $\omega a/v_p=1/\sqrt{12}$ for the $A_0$ mode and $\omega a/v_p=1$ for the $S_0$ one. As can be seen, the device works properly for the two modes, being the focusing point identical in both situations. It can also be seen that the field distribution is not the same for the two modes, given that, although the $A_0$ mode can be scattered almost without interaction with other Lamb modes, the $S_0$ mode tends to excite $SH_0$ waves, hence the displacement field is not the same. In other words, this theory provides the refractive index for the two modes, but not the mode's dynamics. Simulations in the mid panels, where the thickness variation has been removed and the plate has a constant thickness $h_b$, show that although the $S_0$ mode is not affected (as expected) the $A_0$ mode it is, given that the field distribution is different now. 


Figure \ref{Fig:field_knumber}, lower panel, shows the normalized field distribution along the central axis of the lens and for the four situations described before. It is clear that the field profile and focusing point is nearly the same for the two modes when the thickness of the plate is position-dependent, but when the thickness is kept constant the $A_0$ mode is different, which clearly verifies the role of the thickness variation.

\begin{figure}
\centering
\includegraphics[width=\columnwidth]{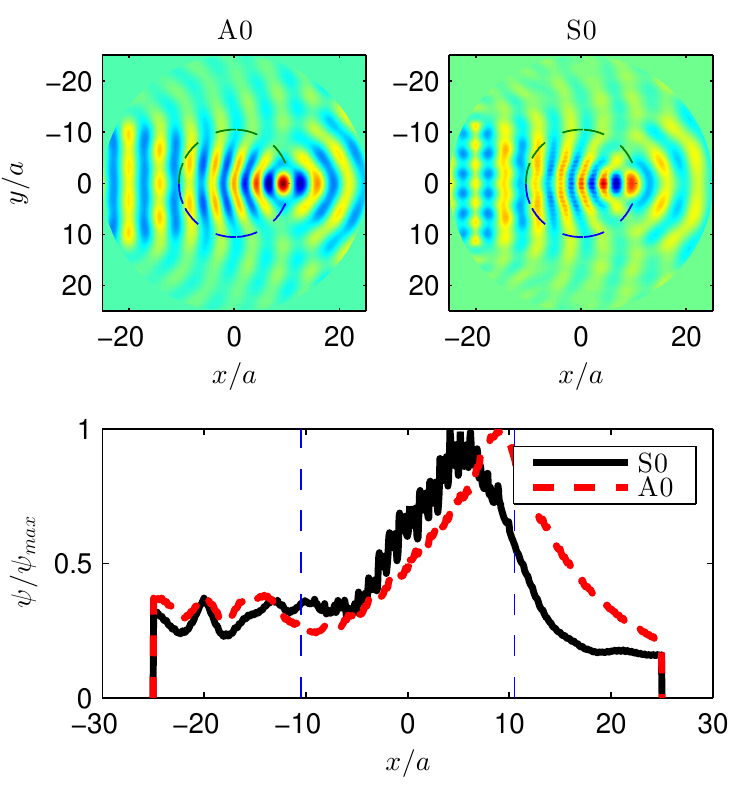}
\caption{\label{Fig:LN} Upper panel: Real part of the field distribution for the $A_0$ (left panel) and $S_0$ (right panel) polarizations interacting with a Luneburg lens designed by means of phononoic crystal with a thickness variation. Lower panel: Absolute value of the field distribution along the $y=0$ line. Vertical dashed lines show the radius of the lens.}
\end{figure}

The other device proposed in the present work is the Luneburg lens. Figure \ref{Fig:LN} shows the performance of a Luneburg lens of radius $R_L=10a$, designed in the same way as the GRIN lens to work for the two modes (inclusions are now arranged in a triangular lattice, which is more adequate for circular clusters). The upper-left panel shows the real part of the normalized $z$ component for the $A_0$ polarization and the right panel shows the real part of the normalized $x$ component for the $S_0$ polarization. The lower panel shows the absolute value of these fields along the axis of the lens. As before the wavelength for the two fields is $ka=1$. In all the plots it can be seen how a focusing point near the border of the lens occurs, as expected.

\begin{figure}
\centering
\includegraphics[width=\columnwidth]{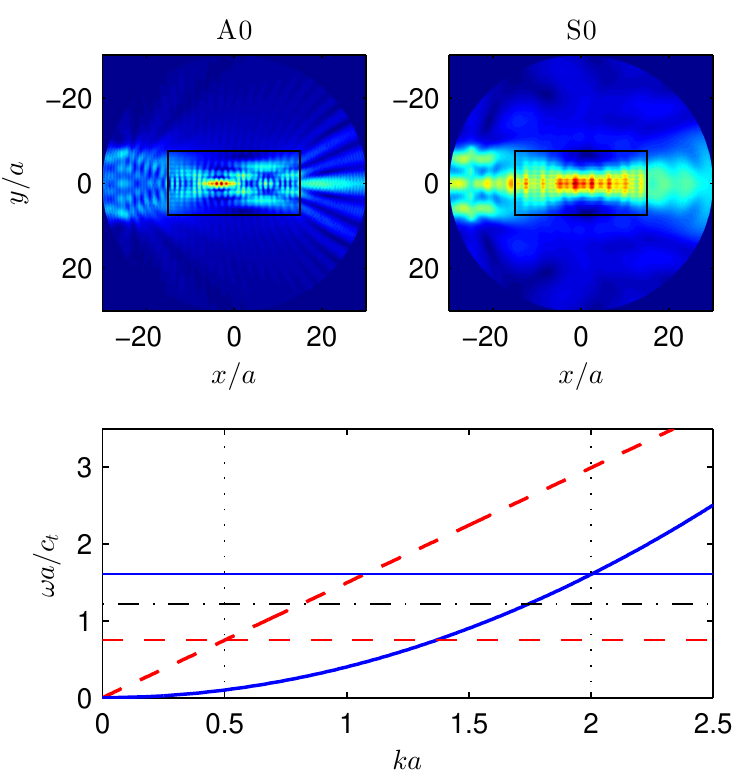}
\caption{\label{Fig:field_omega} Field distributions for the interaction of $A_0$ mode (upper-left panel) and the $S_0$ mode (upper-right panel) working at the same frequency but different wavenumbers. Lower panel: Dispersion relation $\omega-k$ for the $A_0$ (blue-continuous line) and the $S_0$ (red-dashed line). Vertical lines show the wavenumber region in which the GRIN lens would work properly, horizontal red-dashed and blue-continuous lines show the frequency region in which the device will work properly for the two modes. The horizontal dot-dashed line shows the frequency at which the simulations in the upper panels have been performed.}
\end{figure}

In the previous simulations it has been shown the performance of the two devices when working at the same wavenumber, but different frequencies, for the two modes. However, for a practical application it is necessary to see the performance of the lens at the same frequency. Figure \ref{Fig:field_omega} shows a COMSOL simulation of the GRIN lens working at the frequency $\omega a/c_t=1.22$, left panel for the $A_0$ mode and right panel for the $S_0$ one. We can see how the focusing point is nearly the same for the two modes, while it is better defined for the $A_0$ mode where the wavelength is $\lambda\approx 3.6a$, while for the $S_0$ mode this wavelength is $\lambda\approx 7.7a$. Similar results were found also for the Luneburg lens.

It must be pointed out that this type of devices is typically limited by two wavelengths. In the lower limit, it is found that for wavelengths larger than the typical size of the device the diffraction dominates over refraction, and the device's focusing properties are altered. In the upper limit, it is found that for wavelengths shorter than 3 or 4 times the lattice constant, the phononic crystal is no longer homogeneous, and the effective medium model cannot be applied. Given that this wavenumber region is well defined, but that the corresponding frequency region is not the same for the two modes, the frequency region in which the device works properly for the two modes will be shorter than that in which it works individually for each of them. The lower panel of figure \ref{Fig:field_omega} shows the dispersion relation for the $S_0$ mode (red-dashed line) and for the $A_0$ mode (blue-continuous line). The two vertical lines define the wavenumber region in which the performance of the device is expected. The frequency region in which the device will work properly for the two modes is limited below for the mode which has the larger wavelength, that is, the $S_0$ mode, and above for the $A_0$ mode, given that it has the shorter wavelength. These limits are shown in the figure by the red-dashed and blue-continuous horizontal lines, respectively. The horizontal black-dot-dashed line shows the frequency at which the simulations were performed. It is seen that there is still a broadband frequency region in which the device can work properly for the two modes.

In summary, this work shows that simultaneous control of the $S_0$ and $A_0$ Lamb modes by means of a graded index phononic crystal, combined with a properly designed thickness variation of the plate, is possible. This control has been done here by designing a GRIN flat lens and a Luneburg lens, and it has been shown a good performance of these devices when interacting with both a $S_0$ polarized wave and a $A_0$ one. These devices work properly in a broadband frequency region for the two modes, so that it is in this frequency region where most of the applications are expected. Other gradient index devices are easily envisioned, and also its extension to the control of the $SH_0$ mode or other type of guided modes is likely. This work opens the door to new type of broadband multimodal devices with a wide variety of applications in all scales.

This work was supported by the ``Agence Nationale de la Recherche (ANR)'' and the ``D\'el\'egation G\'en\'erale a l’Armement (DGA)'' under the project Metactif, Grant No. ANR-11-ASTR-015. Yabin Jin acknowledges a scholarship provided by China Scholarship Council No.201406260170.

%

\end{document}